Use LATEX !

-------------------------------CUT HERE
----------------------------------------------------

\documentstyle{article}
\begin{document}
\baselineskip = 14pt

\title{\huge \bf R-matrices for the semicyclic
representations of $ U_q\widehat {sl}(2) $ }
\author {I.T.Ivanov\footnotemark \hspace{.2in}  and \hspace{.2in} D.B.Uglov \\
Department of Physics, State University of New York
at Stony Brook\\ Stony Brook, NY 11794-3800}
\date{9 March 1992}
\maketitle

\begin{abstract}
 R-matrices for the semicyclic representations of $ U_q\widehat{sl}(2) $ are
found as a
specialization in the R-matrices of the checkerboard chiral Potts model \\
\end{abstract}

\footnotetext{e-mail iti@dirac.physics.sunysb.edu}

 The chiral Potts model \cite{Potts}\cite{A128}  is the first example of a new
class
of integrable statistical mechanical  models whose Boltzmann weights are
meromorphic functions on an algebraic variety of genus greater than 1.
{}From the mathematical point of view this model is related to the theory of
cyclic representations of the quantum algebra
$ U_q\widehat{sl}(2) $ at $q^N=1$ \cite{bazh}.
This relation is expressed by the fact that the R-matrix intertwining two
tensor products of cyclic representations is built up of the Boltzmann weights
of the chiral Potts model \cite{bazh}\cite{jim}.

In this paper we consider certain limit of the chiral Potts model which
in the algebraic language corresponds to the descent from the cyclic
representation to the semicyclic one \cite{rep}. Boltzmann weights remain well
defined in this limit and as a result we obtain  R-matrices which intertwine
semicyclic representations of  $U_q \widehat{sl}(2)$ and satisfy the
Yang-Baxter equation.

  Series of R-matrices were
obtained for semicyclic representations of $U_qsl(2)$ in
\cite{spain}\cite{spainN}.
The R-matrices obtained here for the larger algebra $U_q\widehat{sl}(2)$ may be
specialized to $U_qsl(2)$ subalgebra.
We checked that in the case $N=3$ the R-matrix from \cite{spain} is equivalent
by
a similarity transform to the R-matrix obtained as a specialization in the
chiral
Potts model in this letter. \\

\paragraph{1.  Checkerboard chiral Potts R-matrix }

         The chiral Potts model Boltzmann weights
\begin{equation}
  \frac{{W}_{\xi,\eta}(i)}{W_{\xi,\eta}(i-1)} =
\frac{\mu_{\xi}}{\mu_{\eta}}\frac
{y_{\eta} - x_{\xi}\omega^{i}}{y_{\xi} - x_{\eta}\omega^{i}}, \quad
 \frac{\overline{W}_{\xi,\eta}(i)}{\overline{W}_{\xi,\eta}(i-1)} = \mu_{\xi}
\mu_{\eta}
 \frac{x_{\xi} \omega - x_{\eta}\omega^{i}}{y_{\eta} - y_{\xi}\omega^{i}}
\end{equation}
depend on two points $\xi$ and $\eta$ of two identical algebraic curves
$C_{\xi}$ and
$C_{\eta}$ defined by the relations
\begin{equation}
\mu^N = \frac{k'}{1-kx^N} = \frac{1-ky^N}{k'}
\end{equation}
and the same for $C_{\eta}$. Also $k^2+k'^2=1$.

  We shall associate \cite{jim} with every cyclic representation $\pi_\xi$ of
the quantum algebra
$U_q\widehat{sl}(2)$ two copies of the curve (2) which we shall denote by
$C_{\xi}$ and $C_{\xi'}$.
The variables of the sets $\xi = {x,y,\mu}$ and $\xi' = {x',y',\mu'}$ as well
as $k$ are
functions of the parameters of the representation . Then the R-matrix
intertwining tensor products of two cyclic representations $\pi_{\xi}$ and
$\pi_{\eta}$ is
given by \cite{jim} :
\begin{eqnarray}
  R(\xi,\eta) = \sum_{i,j,k,l = 1}^{N}
\omega^{i(j-k)}\left( Z^{j} \otimes Z^{k} \right) \left( X^{-1} \otimes X
\right)^{i+l}
\times\nonumber \\
\overline{W}_{\xi,\eta}(k) \overline{W}_{\xi',\eta'}(j)
\widehat{W}_{\xi',\eta}(i) \widehat{W}_{\xi,\eta'}(l)
\end{eqnarray}
\newcommand{\e}{\varepsilon}
where $Z$ and $X$ are $N$ by $N$ matrices satisfying :
\begin{eqnarray}
ZX = \omega XZ \quad Z^N = X^N = 1 \quad \omega = q^2
\end{eqnarray}
and $\widehat{W}_{\xi,\eta}$ is a Fourier transform of the Boltzmann weight (1)
:
$\widehat{W}_{\xi,\eta}(i) = \sum_{k=1}^N W_{\xi,\eta}(k) \omega^{-ik}$.
\begin{equation}
\frac{\widehat{W}_{\xi,\eta}(i)}{\widehat{W}_{\xi,\eta}(i-1)} =
\frac{\mu_{\eta} y_{\xi} - \mu_{\xi} y_{\eta} \omega^{i-1}}{\mu_{\eta} x_{\eta}
 - \mu_{\xi} x_{\xi} \omega^i} \nonumber
\end{equation}
    Throughout this paper we consider only the case N odd .\\
    Matrix elements of
the R-matrix (3) are vertex weights of the checkerboard  chiral Potts model
\cite{A128},
\cite{checker} \\

\paragraph{2. Taking limit in the Boltzmann weights.}

  Here we take a special limit in the checkerboard chiral Potts model which
will allow us
to get the R-matrices for the semicyclic representations.
      It is shown in the appendix A that taking the semicyclic limit in the
cyclic representation
is equivalent to the following rescaling of the variables :
\begin{eqnarray}
 k = \e^{-N} K  ,\quad        k' = \e^{-N} K' \\
x = \e X ,\quad y = \e^{-1} Y ,\quad \mu = \e^{-1} M ,\quad x' = \e^{-1} X'
,\quad
y' = \e Y' ,\quad \mu' = \e M'
\end{eqnarray}
and subsequently setting $\e$ to zero.
       The Boltzmann weights entering the R-matrix are invariant under this
rescaling :
\begin{equation}
 \frac{W_{\xi',\eta}(i)}{W_{\xi',\eta}(i-1)} = \frac{M_{\xi'}}{M_{\eta}}
\frac{Y_{\eta} -
X_{\xi'}\omega^{i}}{Y_{\xi'} - X_{\eta} \omega^{i}} \qquad
 \frac{W_{\xi,\eta'}(i)}{W_{\xi,\eta'}(i-1)} = \frac{M_{\xi}}{M_{\eta'}}
\frac{Y_{\eta'} -
X_{\xi}\omega^{i}}{Y_{\xi} - X_{\eta'} \omega^{i}}
\end{equation}
\begin{equation}
 \frac{\overline{W}_{\xi,\eta}(i)}{\overline{W}_{\xi,\eta}(i-1)} = M_{\xi}
M_{\eta}
\frac{X_{\xi}\omega - X_{\eta}\omega^{i}}{Y_{\eta} - Y_{\xi}\omega^{i}} \qquad
 \frac{\overline{W}_{\xi',\eta'}(i)}{\overline{W}_{\xi',\eta'}(i-1)} = M_{\xi'}
M_{\eta'}
\frac{X_{\xi'}\omega - X_{\eta'}\omega^{i}}{Y_{\eta'} - Y_{\xi'}\omega^{i}}
\qquad
\end{equation}
     The finite parts of the parameters $X,\,Y \ldots$ obey the following
equations :
\begin{equation}
 M^N = \frac{k'}{1-K X^N} = \frac{\e^{2N} - K Y^N}{K'} \qquad
 {M'}^N = \frac{k'}{\e^{2N} - K {X'}^N} = \frac{1 - K{Y'}^N}{K'}
\end{equation}
with $K^2 + {K'}^2 = {\e}^{2N}$.
     Let us now take the limit $\e \rightarrow 0$. The Boltzmann weights (7-8)
remain well
defined.
    The equations of the curve show that $M$ is proportional to $Y$ and
 $M'$ is prpoprional to ${1}/{X'}$.
So the Boltzmann weights (8,9) simplify and to describe them more neatly
 we introduce one more set of parameters
${A,B,C,D}$ defined by
\begin{equation}
A=K^{1/N} Y^{-1} \quad B = - K^{1/N} X \quad C = K^{1/N} X'^{-1} \quad D = -
K^{1/N} Y'
\end{equation}
  This set of parameters satisfy the equations of two Fermat curves :
\begin{equation}
  A^N = B^N + 1 \qquad C^N = D^N + 1
\end{equation}
 The Boltzmann weights entering the R-matrix finally have the following form :

\begin{eqnarray}
\frac{{W}_{\xi,\eta'}(i)}{W_{\xi,\eta'}(i-1)} = \frac{D_{\eta} - B_{\xi}
\omega^i}
  {C_{\eta} - A_{\xi}\omega^{i}} \quad
\frac{{W}_{\xi',\eta}(i)}{W_{\xi',\eta}(i-1)} = \frac{C_{\xi} - A_{\eta}
\omega^{i}}
  {D_{\xi} - B_{\eta}\omega^{i}}
\end{eqnarray}
\begin{eqnarray}
\frac{\overline{W}_{\xi,\eta}(i)}{\overline{W}_{\xi,\eta}(i-1)} =
\frac{B_{\xi}\omega -
  B_{\eta}\omega^i}{A_{\xi} - A_{\eta}\omega^{i}} \quad
\frac{\overline{W}_{\xi',\eta'}(i)}{\overline{W}_{\xi',\eta'}(i-1)} =
\frac{C_{\eta}\omega -
C_{\xi}\omega^i}{D_{\eta} - D_{\xi}\omega^{i}}
\end{eqnarray}

The periodicity property $W(i+N) = W(i),\,\overline{W}(i+N) = \overline{W}(i)$
is
guaranteed by the equations (12).

\paragraph{3. R-matrix for the semicyclic representations of
$U_q\widehat{sl}(2)$}
Here we present the R-matrix and associated Hamiltonian.\\
 The R-matrix intertwining two semicyclic representations $\pi_{\xi}$ and
$\pi_{
\eta}$ (See Appendix A) as a result of the previous section is also given by
the formula
(3) with Boltzmann weights (13,14).
The intertwining condition can be proved along the same lines as in the
periodic  case \cite{jim}.
 To prove the
Yang-Baxter equation we introduce four auxiliary Boltzmann weights :
\begin{eqnarray}
\frac{W_{\xi,\eta}(i)}{W_{\xi,\eta}(i-1)} = \frac{1 - \phi B_{\xi} A_{\eta}
\omega^i}
   {1 - \phi A_{\xi} B_{\eta} \omega^i} \quad
\frac{W_{\xi',\eta'}(i)}{W_{\xi',\eta'}(i-1)} = \frac{\phi C_{\xi} D_{\eta} -
\omega^i}
   {\phi D_{\xi} C_{\eta} - \omega^i}   \\
\frac{\overline{W}_{\xi',\eta}(i)}{\overline{W}_{\xi',\eta}(i-1)} =
\frac{\omega -
  \phi C_{\xi} B_{\eta} \omega^i}{1 - \phi D_{\xi} A_{\eta} \omega^i}  \quad
\frac{\overline{W}_{\xi,\eta'}(i)}{\overline{W}_{\xi,\eta'}(i-1)} = \frac{\phi
B_{\xi}
 C_{\eta} \omega - \omega^i}{\phi A_{\xi} D_{\eta} - \omega^i}
\end{eqnarray}
    We checked that the Boltzmann weights (13-14),(15-16) satisfy a set of
eight
star-tringle equations
\begin{equation}
\sum_{d=1}^N \overline{W}_{q,r}(b-d) W_{p,r}(a-d) \overline{W}_{p,q}(d-c) =
R_{p,q,r} W_{p,q}(a-b) \overline{W}_{p,r}(b-c) W_{q,r}(a-c)
\end{equation}
with $p$,$q$ and $r$ replaced by both primed and unprimed indices and $\phi$
arbitrary number.
This set of equations imply the Yang-Baxter equation for the semicyclic
R-matrix of the
form (3).The fact that the R-matrix (3) satisfies the Yang-Baxter equation can
also be derived from the intertwining property and an indecomposability of the
third-order tensor product of the semicyclic representations of
$U_q\widehat{sl}(2)$ at generic values of the parameters.
If the restrictions  $D=-qA,B=-q^{-1}C$
are imposed on the parameters $A,B,C,D$ we obtain $e_0=f_1$, and $f_0=e_1$
in the representation of $U_q\widehat{sl}(2)$ (see Appendix A) .And the
resulting R-matrix intertwins between
the representations of finite dimensional quantized algebra $U_q sl(2)$.
In the case $N=3$ we checked that this R-matrix is equivalent
to the R-matrix of \cite{spain} by the similarity transformation.

The spin-chain  hamiltonian corresponding to the R-matrix (3) can be obtained
using the standard procedure \cite{A128}.To write down it we introduce the
following operators: \begin{equation}
X_i= \stackrel{1}{1}\otimes \cdots \otimes \stackrel{i}{X} \otimes \cdots
\otimes \stackrel{L}{1}
\quad Z_i= \stackrel{1}{1}\otimes \cdots \otimes \stackrel{i}{Z} \otimes \cdots
\otimes \stackrel{L}{1}
\end{equation}
where $i=1,\cdots,L$ ; L is the length of the periodic lattice , X  and Z are
the same as in (4).In terms of these operators the hamiltonian reads as
follows:
\begin{equation}
H=\sum_{j=1}^{L} h_{j,j+1},\\
\end{equation}
\begin{equation}
h_{j,j+1}= \sum_{n=1}^{N} F(n) X_j^{-n}X_{j+1}^n + \sum_{n,m=1}^{N}
[ G_+(n,m) X_j^{-n}X_{j+1}^n Z_j^m + G_-(n,m) X_j^{-n}X_{j+1}^n Z_{j+1}^m  ]
\end{equation}
where the explicit form of the coefficients $F,G_+,G_-$ is given in the
appendix B.
The form of this hamiltonian is the same as of the general {\it checkerboard}
chiral Potts model. In the case of the well-known chiral Potts model the
coefficients $G_+(n,m)$ and $G_-(n,m)$ become proportional to $\delta_{n,0}$
and the hamiltonian acquires the usual form \cite{Potts}.\\

\appendix{\it{Appendix A}}\\
In this appendix we describe the $U_q\widehat{sl}(2)$ algebra and its cyclic
and semicyclic representations at  $q^N=1$. After that we describe in detail
the limiting procedure used in section 2.

   Quantized enveloping algebra $U_q\widehat{sl}(2)$ is generated by the
elements $t_i , t_i^{-1} ,
 e_i$ and $f_i \; (i=0,1)$ subject to the relations
\begin{eqnarray}
t_i t_i^{-1} = t_i^{-1} t_i = 1 \quad t_i t_j = t_j t_i  \\
t_i e_j t_i^{-1} = q^{a_{ij}} e_j ,\quad t_i f_j t_i^{-1} = q^{-a_{ij}} f_j,
\quad
[ e_i , f_j ] = \frac{t_i - t_i^{-1}}{q_i - q_i^{-1}} \delta_{ij}
\end{eqnarray}
with $a_{ij} = 2, i=j \> and \> a_{ij} = -2 , i\neq j $. These generators also
obey the
Serre relations.

   Tensor products of representations are formed using the following
comultiplication
\begin{eqnarray}
\Delta(e_0) = e_0 \otimes 1 + z^{-1}t_0 \otimes e_0 , \quad
\Delta(e_1) = e_1 \otimes 1 + z \, t_1 \otimes e_1 \\
\Delta(f_0) = f_0 \otimes t_0^{-1} + z \otimes f_0 , \quad
\Delta(f_1) = f_1 \otimes t_1^{-1} + z^{-1} \otimes f_1 \\
\Delta(t_i) = t_i \otimes t_i , \quad \Delta(z) = z \otimes z
\end{eqnarray}
$z$ is central element commuting with the algebra $U_q\widehat{sl}(2)$.
    N-dimensional cyclic representation of the quantum algebra
$U_q\widehat{sl}(2)$ may be
described in terms of the two matrices $Z$ and $X$
and five complex parameters $\xi = (a,b,x_0,x_1,c)$,
\begin{eqnarray}
\pi_\xi(e_0) = x_0 X^{-1} \frac{b^2 Z^{-1} - 1}{q - q^{-1}} ,\quad
\pi_\xi(f_0) = (a b x_0)^{-1} \frac{a^2 Z - 1}{q-q^{-1}}X \\
\pi_\xi(f_1) = (a b x_1)^{-1} X^{-1} \frac{b^2 Z^{-1} - 1}{q-q^{-1}},\quad
\pi_\xi(e_1) = x_1 \frac{a^2 Z - 1}{q-q^{-1}} X \\
\pi_\xi(t_0) = \frac{b Z^{-1}}{q a} ,\quad \pi_\xi(t_1) = \frac{q a Z}{b}
 ,\quad \pi_\xi(z) = c
\end{eqnarray}

   R-matrix is an operator which intertwins tensor products of two
representations $\pi_\xi$
and $\pi_\eta$ taken in an alternated order :
\begin{equation}
R(\xi,\eta) (\pi_\xi \otimes \pi_\eta) \Delta (g) = (\pi_\eta \otimes \pi_\xi)
\Delta(g)
R(\xi,\eta) \;,\; for \; all\, g\, \in U_q\widehat{sl}(2).
\end{equation}
     Necessary and sufficient condititon for an intertwiner between tensor
products of two
cyclic representations $\pi_{\xi}$ and $\pi_{\eta}$ to exist is that the
following four
quantities should be the same in both representations.
\begin{eqnarray}
\gamma_1 = \frac{x_1^N(a^{2N}-1)}{1-a^N b^{-N} c^N}, \qquad
\gamma_2 = \frac{a^{2N} -1}{a^N b^N x_0^N (a^N b^{-N} - c^N)}  \\
\gamma_3 = \frac{b^{2N} -1}{a^N b^N x_1^N (a^{-N} b^{N} - c^{-N})}, \qquad
\gamma_4 = \frac{x_0^N (b^{2N} - 1)}{1 - a^{-N} b^{N} c^{-N}}
\end{eqnarray}
   These four invariants are not independent but obey $\gamma_1 \gamma_3 =
 \gamma_2 \gamma_4$.
   In order to write the R-matrix for the cyclic case in the factorized form a
change of parameters $\{\gamma_1,\gamma_2,\gamma_3,\gamma_4,a,b,c,x_0,x_1\}
\mapsto \{k,k_0,k_1,x,y,\mu,x',y',\mu'\}$ is convenient.
\begin{eqnarray}
  k^2 = \frac{-1}{\gamma_1 \gamma_3}, \quad k_0^N = - k \gamma_4, \quad k_1^N =
- k
 \gamma_1 \\
  x_0 = k_o x',\quad x_1 = k_1 y',\quad a^2 = \frac{x \mu \mu'}{y'},\quad
b^2 = \frac{y}{x' \mu \mu'},\quad c = a b \frac{y'}{q x}
\end{eqnarray}
     Then the R-matrix satisfying (29) is given by (3) with the chiral Potts
Boltzmann weights.

 We will call N-dimensional representation of $U_q\widehat{sl}(2)$ semicyclic
if $e_1^{N}$ and
$f_0^{N}$ are represented by non-zero scalars while $e_0^{N}=f_1^{N}=0$ . In
terms of the
parameters of the cyclic representation (26-28) this means $b^{2 N} - 1 = 0$.\\
Let $\pi_{\xi} \,
\xi = (a,b,c,x_0,x_1)$ and $\pi_{\eta} \, \eta=
(\hat{a},\hat{b},\hat{c},\hat{x}_0,\hat{x}_1)$
are two cyclic representations. Let also

\begin{equation}
     b^{2 N} - 1 = \e^{2 N} f  \qquad  \hat{b}^{2 N} - 1 = \e^{2 N} \hat{f}
\end{equation}
where $f$ and $\hat{f}$ are some finite functions of the parameters of the
corresponding
representations and $\e$ is a number. In limit the $\e \rightarrow 0$ both
representations
$\pi_{\xi}$ and $\pi_{\eta}$ become semicyclic. Let us using the equations
(32-33) to explicitely
single  out powers of $\e$ from the invariants and representation parameters :
\begin{eqnarray}
\gamma_3 = \e^{2N} \Gamma_3 =\e^{2N} f G_3,\quad \gamma_4 = \e^{2N} \Gamma_4
=\e^{2N} f G_4
\quad \gamma_1 = \Gamma_1 ,\quad \gamma_2 = \Gamma_2  \\
 k = \e^{-N} K ,\quad k_0 = \e K_0 ,\quad k_1 = \e^{-1} K_1 \\
x = \e X ,\quad y = \e^{-1} Y ,\quad \mu = \e^{-1} M ,\quad x' = \e^{-1} X'
,\quad
y' = \e Y' ,\quad \mu' = \e M'
\end{eqnarray}
     As is mentioned in the main text the Boltzmann weights are invariant under
this
rescaling and we may safely set $\e$ to zero.
     The invariants $\Gamma_1$ and $\Gamma_2$ still play the role of necessary
conditions
for existing of the intertwiner. The initial invariant conditions for
$\gamma_3$ and
$\gamma_4$ become an defining equation for the functions $f$ and $\hat{f}$ from
(34):
\begin{equation}
 \frac{f}{\hat{f}} = \frac{\hat{G}_3}{G_3} = \frac{\hat{G}_4}{G_4}
\end{equation}
  The Boltzmann weights entering R-matrix (3) do not depend on the particular
choice of
$f$ and $\hat{f}$
satisfying (38). The four other possible Boltzmann weights become trivial in
the
semicyclic limit :
\begin{equation}
 \frac{W_{\xi,\eta}(i)}{W_{\xi,\eta}(i-1)} = 1 \quad \frac{W_{\xi',\eta'}(i)}
{W_{\xi',\eta'}(i-1)} = 1  \quad \frac{\overline{W}_{\xi',\eta}(i)}
{\overline{W}_{\xi',\eta}(i-1)} = \omega \quad
\frac{\overline{W}_{\xi,\eta'}(i)}
{\overline{W}_{\xi,\eta'}(i-1)} = 1 \quad
\end{equation}

\appendix{\it{Appendix B}}

Here we list the coefficients $G_+,G_-,F$ entering the hamiltonian (20):

\begin{eqnarray}
F(n)= \sum_{i= 1}^{N} \widehat{W}_{\xi,\xi^{'}}(n-i)
\widehat{W}_{\xi^{'},\xi}(i)
[ \sum_{s=1}^{n-i} \frac{\gamma-\delta \omega^{s-1}}
{C_\xi-D_\xi\omega^s} - \sum_{s=1}^{i}\frac{\beta-\alpha\omega^{s-1}}
{B_\xi-A_\xi\omega^s}]   ;\\
G_+(n,m)=E(n,m)\overline{W}_{\xi^{'},\xi^{'}}(m)  ; \quad
G_-(n,m)=E(n,-m)\overline{W}_{\xi,\xi}(m)
\end{eqnarray}
where
\begin{eqnarray}
E(n,m)=\sum_{i= 1}^{N} \omega^{i m}  \widehat{W}_{\xi,\xi^{'}}(i)
\widehat{W}_{\xi^{'},\xi}(n-i) ;\\
\end{eqnarray}
and
\begin{eqnarray}
\overline{W}_{\xi^{'},\xi^{'}}(i)=\frac{\gamma}{C_\xi}
\frac{({\frac{C_\xi}{D_\xi}})^{i}}{1-\omega^i} ;
\overline{W}_{\xi,\xi}(i)=\frac{\beta}{B_\xi}
\frac{({\frac{B_\xi}{A_\xi}})^{i}}{1-\omega^{-i}} ;\\
\widehat{W}_{\xi^{'},\xi}(i)= \prod_{s=1}^{i}
\frac{C_\xi-D_\xi\omega^{s-1}}{A_\xi-B_\xi\omega^s}  ;
\widehat{W}_{\xi,\xi^{'}}(i)= \prod_{s=1}^{i}\frac{D_\xi-C_\xi\omega^{s-1}}
{B_\xi-A_\xi\omega^s}  ;
\end{eqnarray}
where the parameters $ A_{\xi} ,  B_{\xi} ,  C_{\xi} ,  D_{\xi} ,
\alpha,\beta,\gamma,\delta$ satisfy the following equations:
\begin{eqnarray}
A_{\xi}^N=B_{\xi}^N+1 ; \quad C_{\xi}^N=D_{\xi}^N+1;\\
\alpha=(\frac{B_{\xi}}{A_{\xi}})^{N-1}\beta;
\gamma=(\frac{D_{\xi}}{C_{\xi}})^{N-1}\delta;
\end{eqnarray}

    {\it Summary}. Quantum enveloping algebras at roots of unity have hierarchy
of
irreducible multiparameter representations : cyclic $\rightarrow$ semicyclic
$\rightarrow$
nilpotent representations. In this letter we have shown that the process of
specializing the
parameters in the cyclic representation of $U_q\widehat{sl}(2)$ to make it
semicyclic goes
through on the R-matrix level also. We believe that the R-matrices from
\cite{spainN} are
equivalent to the R-matrices obtained here when the later are specialized to
the $U_qsl(2)$
case. The limit to the nilpotent representations \cite{berk} is under
investigation.

The R-matrices for the cyclic representations of $U_q\widehat{sl}(2)$ are
already known
\cite{sln} and it seems feasible to find the semicyclic limit in this case.\\

    {\it Acknowdgement} \\
We  are grateful to prof. B.M.McCoy for many helpful suggestions   and to
R.Kedem for
discussions.

\end{document}